\documentclass[aps,prl,preprint,groupedaddress]{revtex4}
\usepackage{graphicx}
\usepackage{amssymb,amsfonts,amsmath}
\usepackage{mathrsfs}
\usepackage{color}
\usepackage[3D]{movie15}
\usepackage{hyperref}
\usepackage{bm}
\usepackage{ulem}
\MakeRobust{\overrightarrow}

\usepackage{epstopdf}
\newif\ifpdf
\ifx\pdfoutput\undefined
   \pdffalse
\else
   \pdfoutput=1
   \pdftrue
\fi
\ifpdf
   \usepackage{graphicx}
   \usepackage{epstopdf}
\else
   \usepackage{graphicx}
\fi

\begin{document}
\title{Geometric Mechanics of Periodic Pleated Origami}
\author{Z.Y.\ Wei $^{1}$,  Z.V. \ Guo $^{1}$, L.\ Dudte $^{1}$, H.Y.\ Liang $^{1}$, L.\ Mahadevan $^{1,2}$}
\affiliation{$^1$ School of Engineering and Applied Sciences,Harvard University, Cambridge, Massachusetts 02138.\\
$^2$ Department of Physics, Harvard University, Cambridge, Massachusetts 02138.\\}

\begin{abstract}
Origami is the archetype of a structural material with unusual mechanical properties that arise almost exclusively from the geometry of its constituent folds and forms the basis for mechanical metamaterials with an extreme deformation response.  Here we consider a simple periodically folded structure Miura-ori, which is composed of identical unit cells of mountain and valley folds with four-coordinated ridges, defined completely by $2$ angles and $2$ lengths. We use the geometrical properties of a Miura-ori plate to characterize its elastic response to planar and non-planar piece-wise isometric deformations and calculate the two-dimensional stretching and bending response of a Miura-ori sheet, and show that the in-plane and out-of-plane Poisson's ratios are equal in magnitude, but opposite in sign. Our geometric approach also allows us to solve the inverse design problem of determining the geometric parameters that achieve the optimal geometric and mechanical response of such structures.
\end{abstract}
\maketitle

Folded and pleated structures arise in a variety of natural systems including insect wings \cite{Forbes1924}, leaves \cite{Kobayashi1998}, flower petals \cite{Kobayashi2003}, and have also been creatively used by origami artists for aeons \cite{Lang}. More recently, the presence of re-entrant creases in these systems that allows the entire structure to fold and unfold simultaneously have also been used in deployable structures such as solar sails and foldable maps \cite{Miura1980, Miura1985, Elsayed2004}.  Complementing these studies, there has been a surge of interest in the mathematical properties of these folded structures \cite{Lang, Demaine, Hull}, and some recent qualitative studies on the physical aspects of origami \cite{Klettand2011, Schenk2011, Alessandro2008}.     In addition, the ability to create them de-novo without a folding template, as a self-organized buckling pattern when a stiff skin resting on a soft foundation is subject to biaxial compression \cite{Whitesides1998, Maha2005, Audoly2008} has opened up a range of questions associated with their assembly in space and time, and their properties as unusual materials.

Here, we quantify the properties of  origami-based 3-dimensional periodically pleated or folded structures, focusing on what is perhaps the simplest of these periodically pleated structure, the Miura-ori pattern (Fig.\ref{Figure1}a) which is defined completely in terms of $2$ angles and $2$ lengths. The geometry of its unit cell embodies the basic element in all nontrivial pleated structures - the mountain or valley fold, wherein four edges (folds) come together at a single vertex,
as shown in Fig.\ref{Figure1}d. It is parameterized by two dihedral angles $\theta\in[0,\pi]$, $\beta\in[0,\pi]$, and one oblique angle $\alpha$, in a cell of length $l$, width $w$, and height $h$.  We treat the structure as being made of identical periodic rigid skew plaquettes joined by elastic hinges at the ridges. The structure can deploy uniformly in the plane (Fig.\ref{Figure1}b) by having each constituent skew plaquette in a unit cell rotate rigidly about the connecting elastic ridges. Then the ridge lengths $l_1$, $l_2$ and $\alpha\in[0,\pi/2]$ are constant through folding/unfolding, so that we may choose $\theta$ (or equivalently $\beta$) to be the only degree of freedom that completely characterizes a Miura-ori cell. The geometry of the unit cell implies that
\begin{align}
\begin{split}
&\beta=2\sin^{-1}(\zeta\sin(\theta/2)),\quad
l=2l_1\zeta,\\
&w=2l_2\xi\quad\text{and}\quad
h=l_1\zeta\tan\alpha\cos(\theta/2),
\end{split}
\label{Figure2geo}
\end{align}
where the dimensionless width and height are
\begin{align}
\begin{split}
\xi=\sin\alpha\sin(\theta/2)\quad \text{and}\quad
\zeta =\cos\alpha(1-\xi^2)^{-1/2}.
\end{split}
\label{xizeta}
\end{align}
We see that $\beta$, $l$, $w$, and $h$ change monotonically as $\theta\in[0,\pi]$, with $\beta\in[0,\pi]$, $l\in 2l_1[\cos\alpha,1]$, $w\in 2l_2[0,\sin\alpha]$, and $h\in l_1[\sin\alpha,0]$. As $\alpha\in[0,\pi/2]$, we see that $\beta\in [\theta, 0]$, $l\in[2l_1, 0]$, $w\in[0,2l_2\sin(\theta/2)]$ and $h\in[0,l_1]$. The geometry of the unit cell implies a number of interesting properties associated with the expansion kinematics of a folded Miura-ori sheet, particularly in the limit of an orthogonally folds when $\alpha=\pi/2$ ({\bf Appendix}; {{\bf{A}}-1), the singular case corresponding to the common map fold where the folds are all independent. More generally, it is possible to optimize the volume of the folded structure as a function of the design variables ({\bf A}-1).

\begin{figure}
\centerline{\includegraphics[width=1\textwidth]{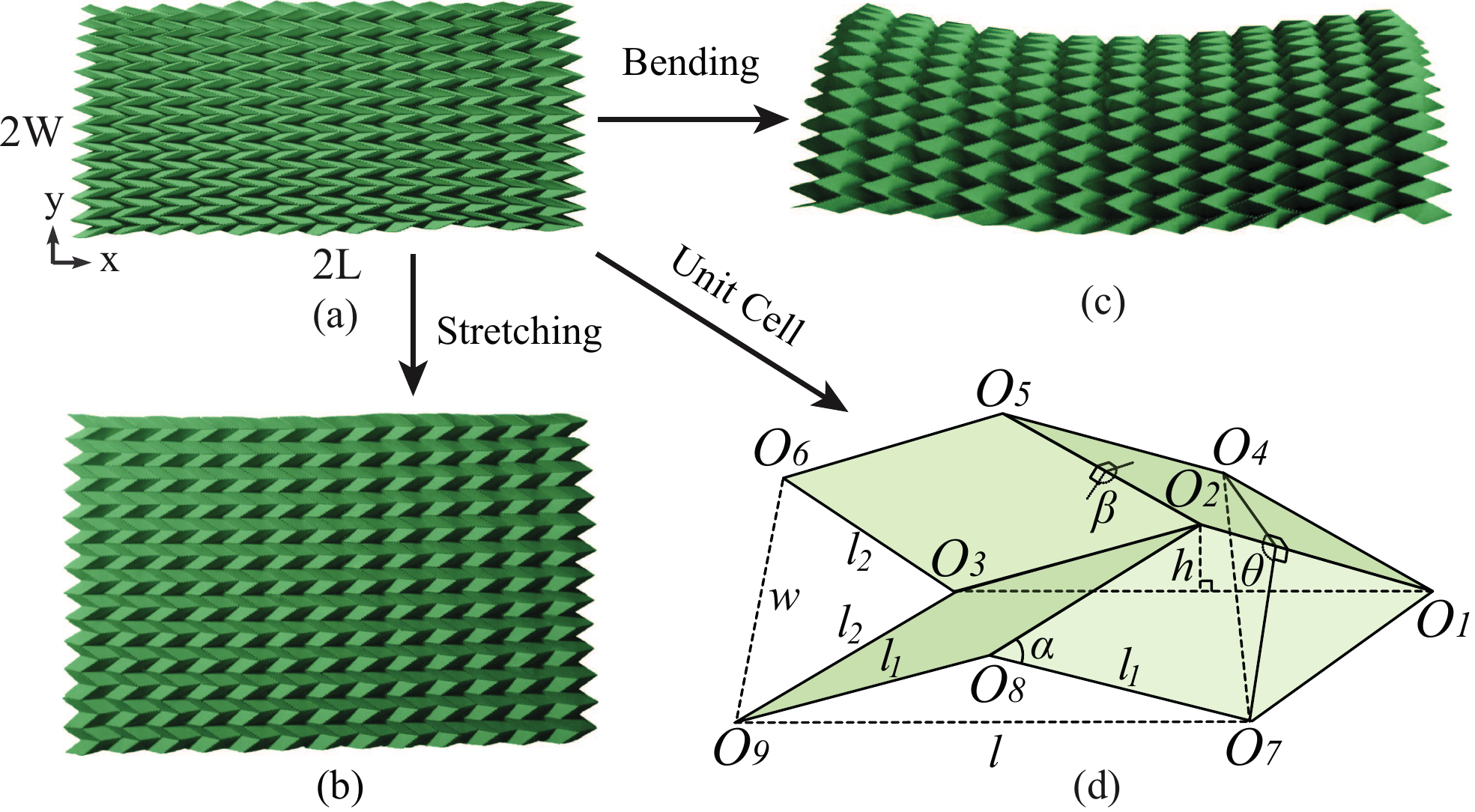}}
\caption{Geometry of Miura-ori pattern. (a) A Miura-ori plate folded from a letter size paper contains $13$ by $13$ unit cells (along $x$ and $y$ direction respectively), with $\alpha=45^o$ and $l_1=l_2=l_e$. The plate dimension is $2L$ by $2W$. (b) In-plane stretching behavior of a Miura-ori plate  when pulled along the $x$ direction shows its expand in all directions, i.e. it has a negative Poisson's ratio. (c) Out-of-plane bending behavior of a Miura-ori plate when a symmetric bending moment is applied on boundaries $x=\pm L$ shows a saddle shape, consistent with that in this mode of deformation its Poisson's ratio is positive. (d) Unit cell of Miura-ori is characterized by two angles $\alpha$ and $\theta$ given $l_1$ and $l_2$ and is symmetric about the  central plane passing through $O_1O_2O_3$.}
\label{Figure1}
\end{figure}

From now on, we assume each plaquette is a rhombus, i.e. $l_1=l_2=l_e$, to keep the size of the algebraic expressions manageable, although it is a relatively straightforward matter to account for variations from this limit. We characterize the planar response of Miura-ori in terms of 2 quantities -- the Poisson's ratio which is a geometric relation that couples deformations in orthogonal directions, and the stretching rigidity which characterizes its planar mechanical stiffness.

The planar Poisson's ratio is defined as
\begin{equation}
\nu_{_{wl}}\equiv-\frac{dw/w}{dl/l} ={1-\xi^{-2}}.
\label{nu_wl}
\end{equation}
The reciprocal Poisson's ratio is $\nu_{_{lw}}=1/\nu_{_{wl}}$. Because $\xi\le1$, the in-plane Poisson's ratio $\nu_{_{wl}}<0$ (Fig.\ref{Figure2}a), i.e. Miura-ori is an auxetic material. To obtain the limits on $\nu_{_{wl}}$, we consider the extreme values of $\alpha, \theta$, since $\nu_{_{wl}}$ monotonically increases in both variables. Expansion of (\ref{nu_wl}) shows that $\nu_{_{wl}}|_{\alpha\rightarrow 0}\sim\alpha^{-2}$, and thus $\nu_{_{wl}}|_\theta\in(-\infty,-\cot^{2}(\theta/2)]$, while $\nu_{_{wl}}|_{\theta\rightarrow0}\sim\theta^{-2}$ and thus $\nu_{_{wl}}|_\alpha\in(-\infty,-\cot^{2}\alpha]$. When $(\alpha,\theta)=(\pi/2,\pi)$, $\nu_{_{wl}}=0$ so that  the two orthogonal planar directions may be folded or unfolded independently when the folds themselves are orthogonal, as in traditional map-folding. Indeed, the fact that this is the unique state for which non-parallel folds are independent makes it all the more surprising that it is still the way in which maps are folded -- since it makes unfolding easy, but folding frustrating! Similar arguments can be applied to determine the other geometric Poisson's ratios related to height changes, $\nu_{_{hl}}$ and $\nu_{_{wh}}$ ({\bf{A}}-2.1).

To calculate the in-plane stiffness of the unit cell, we note that the potential energy of a unit cell deformed by a uniaxial force $f_x$ in the $x$ direction reads, $
H=U-\int_{\theta_0}^{\theta} f_x (dl/d \theta') d\theta'$, where the elastic energy of a unit cell is stored only in the elastic hinges which allow the plaquettes to rotate, with $U=kl_e(\theta-\theta_0)^2+kl_e(\beta-\beta_0)^2$, $k$ being the hinge spring constant, $\theta_0$ and $\beta_0$ $(=\beta(\alpha,\theta_0))$ being the natural dihedral angles in the undeformed state. The external force $f_x$ at equilibrium state is obtained by solving the equation $\delta H/\delta \theta=0$ ({\bf{A}}-2.2), while the stretching rigidity  associated with the $x$ direction is given by
\begin{equation}
K_x(\alpha,\theta_0)\equiv\frac{df_x}{d\theta}\bigg\vert_{ _{\theta_0}}
=\frac{4k[(1-\xi_0^2)^{2}+\cos^2\alpha]}{(1-\xi_0^2)^\frac{1}{2}\cos\alpha\sin^2\alpha\sin\theta_0},
\label{Kx}
\end{equation}
where $\xi_0 = \xi(\alpha,\theta_0)$ and $\xi$ is defined in (\ref{xizeta}). To understand the limits of $K_x$, we expand (\ref{Kx}) in the vicinity of the extreme values of $\alpha$ and $\theta_0$ which gives us $K_x\sim \alpha^{-2}$ as $\alpha\rightarrow0$, $K_x\sim (\pi/2-\alpha)^{-1}$ as $\alpha\rightarrow\pi/2$, $K_x\sim \theta^{-1}$ as $\theta\rightarrow0$, $K_x\sim (\pi-\theta)^{-1}$ as $\theta\rightarrow\pi$. We see that  $K_x$ has a singularity at $(\alpha,\theta) = (\pi/2,\pi)$.

We note that $K_x$ is not monotonic in either $\alpha$ or $\theta_0$, so that there is an optimal pair of these variables for which the stiffness is an extremum.  Setting $\partial_{\theta_0} K_x|_\alpha=0$ and $\partial_\alpha K_x|_{\theta_0}=0$ allows us to determine the optimal design curves, $\theta_{0m}(\alpha)$ (green dotted curve in Fig.\ref{Figure2}b) and $\alpha_m(\theta_0)$ (red dashed curve in Fig.\ref{Figure2}b) that correspond to the minimum value of the stiffness $K_x$ as a function of the underlying geometric parameters defining the unit cell. These curves are monotonic, and furthermore $\theta_{0m}(\alpha)$ is perpendicular to $\alpha=0$, because when $\alpha\rightarrow0$ it is asymptotically approximated by $4(\theta_{0m}-\pi/2)=\alpha^2$ ({\bf{A}}-2.3). Similarly, $\alpha_m(\theta_0)$ is perpendicular to $\theta_0=0$, because when $\theta_0\rightarrow0$ it is asymptotically approximated by $c(\alpha_m-\alpha^*)=\theta_0^2$, where $c = 4\sqrt{5+5\sqrt{5}}$ and $\alpha^*=\cos^{-1}\sqrt{\sqrt{5}-2}\approx 60.9^o$. Analogous arguments allow us to determine the other stretching rigidity $K_y$, which is coupled to $K_x$ through design angles $\alpha$ and $\theta$ ({\bf{A}}-2.2, 2.3).

\begin{figure}
\centerline{\includegraphics[width=1\textwidth]{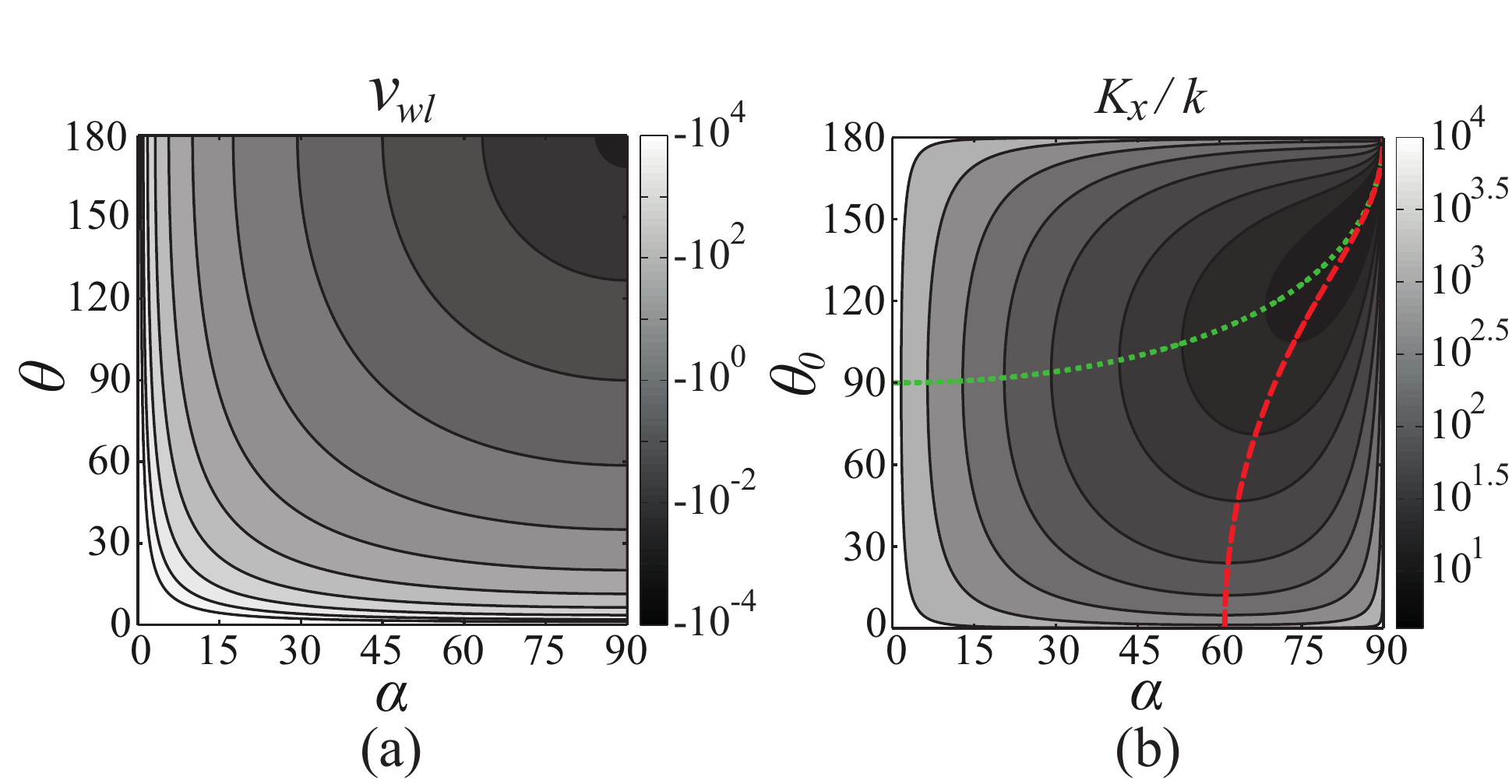}}
\caption{ In-plane stretching response of a unit cell. (a) Contour plot of Poisson's ratio $\nu_{_{wl}}$. $\nu_{_{wl}}$ shows that it monotonically increases with both $\alpha$ and $\theta$. $\nu_{_{wl}}|_\alpha\in[-\infty,-\cot^{2}\alpha]$, and $\nu_{_{wl}}|_\theta\in[-\infty,-\cot^{2}(\theta/2)]$. (b) Contour plot of the dimensionless stretching rigidity $K_x/k$. The green dotted curve indicates the optimal design angle pairs that correspond to the minima of $K_x|_{\alpha}$. The red dashed curve indicates the optimal design angle pairs that correspond to the minima of $K_x|_{\theta_0}$.  See the text for details.}
\label{Figure2}
\end{figure}

To understand the bending response of Miura-ori, we must consider the conditions when it is possible to bend a unit cell isometrically, i.e. with only rotations of the plaquettes about the hinges. Geometric criteria show that planar folding is the only possible motion using {rigid rhombus plaquettes} in our Miura-ori plates ({\bf{A}}-3.1). To enable the bending mode, the minimum model for isometric deformations requires the introduction of $1$ additional diagonal fold into each plaquette (Fig.\ref{Figure3}a), either the short fold (e.g. $O_2O_7$) or the long one (e.g. $O_1O_8$). Here, we adopt the short fold as a result of which $4$ additional DOFs arise and allow both symmetric bending and asymmetric twisting, depending on whether the rotations are symmetric or not.

We see that the out-of-plane bending (Fig.\ref{Figure1}c) has Poisson's ratio $\nu_b\equiv-\kappa_y/\kappa_x>0$ \footnote{In general, the incremental Poisson's ratio is $\nu_b=-d\kappa_y/d\kappa_x$, but here we only consider linear deformation near the rest state, so $\nu_b=-\kappa_y/\kappa_x$}, where $\kappa_x$ and $\kappa_y$ are curvatures in the $x$ and $y$ directions. To calculate $\nu_b$ in linear regime, where the rotations are infinitesimal, we need to first derive the expressions for both curvatures. If $\kappa_x$ is the curvature in the $x$ direction, it may be expressed as the dihedral angle between plane $O_6O_3O_9$ and $O_4O_1O_7$ (Fig.\ref{Figure3}a) projected onto the $x$ direction over the unit cell length. {Similarly, the other curvature component $\kappa_y$ may be expressed as the dihedral angle between plane $O_4O_5O_6$ and $O_7O_8O_9$ projected onto the $y$ direction over the unit cell width. These are given by
\begin{equation}
\begin{split}
\kappa_x &= \frac{\cos(\alpha/2)\sin(\theta/2)}{2l_e\sqrt{1-\xi^2}}(\phi_2+\phi_4),\\
\kappa_y &= -\frac{\sqrt{1-\xi^2}}{4l_e\sin(\alpha/2)\xi }(\phi_2+\phi_4).
\end{split}
\label{kappa}
\end{equation}
where $\phi_2$, $\phi_4$ are rotation angles about internal folds $\overrightarrow{O_7O_2}$ and $\overrightarrow{O_8O_3}$ respectively, which are positive according to the right-hand rule} ({\bf{A}}-3.2). We note that although there are a total of $5$ deformation angles (Fig.\ref{Figure3}a), both $\kappa_x$ and $\kappa_y$ depend only on $\phi_2$ and $\phi_4$. This is because of the symmetry of deformations about $xoz$ plane; $\phi_3$ and $\phi_5$ are functions of $\phi_1$ and $\phi_2$ (Eq. \ref{phis} in {\bf{A}}), and the case that $\phi_1$ changes while keeping $\phi_2$ and $\phi_4$ being 0 corresponds to the planar stretch of a unit cell, so $\phi_1$ does not contribute to both curvatures. This is consistent with our intuition that bending a unit cell requires the bending of plaquettes. The Poisson's ratio for bending is thus given by
\begin{equation}
\nu_b = -\frac{\kappa_y}{\kappa_x}= -1+\xi^{-2} = -\nu_{_{wl}},
\label{nub}
\end{equation}
where the last equality follows from Eqs. (\ref{nu_wl}) and (\ref{kappa}). {If the original plaquettes are allowed to fold along the long diagonals instead (e.g. $O_8O_1$ in Fig.\ref{Figure3}a), the new curvature components $\kappa_x$ and $\kappa_y$ are still given by (\ref{kappa}) with  $\alpha$ being replaced by $\pi-\alpha$ ({\bf A}-3.3), and $\phi_2,\phi_4$ now being rotations about axis $\overrightarrow{O_8O_1}$ and $\overrightarrow{O_9O_2}$ respectively. Therefore $\nu_b=-{\kappa_y}/{\kappa_x}=-\nu_{_{wl}}$.}  This result, that the in-plane Poisson's ratio is negative while the out-of-plane Poisson's ratio is positive, but has the same magnitude is independent of the mechanical properties of the sheet and is a consequence of geometry alone. Although our analysis is limited to the case when the deformation involves only small changes in the angles about their natural values, it is not as restrictive as it seems, since small changes to the unit cell can still lead to large global deformations of the entire sheet.

\begin{figure}
\centerline{\includegraphics[width=1\textwidth]{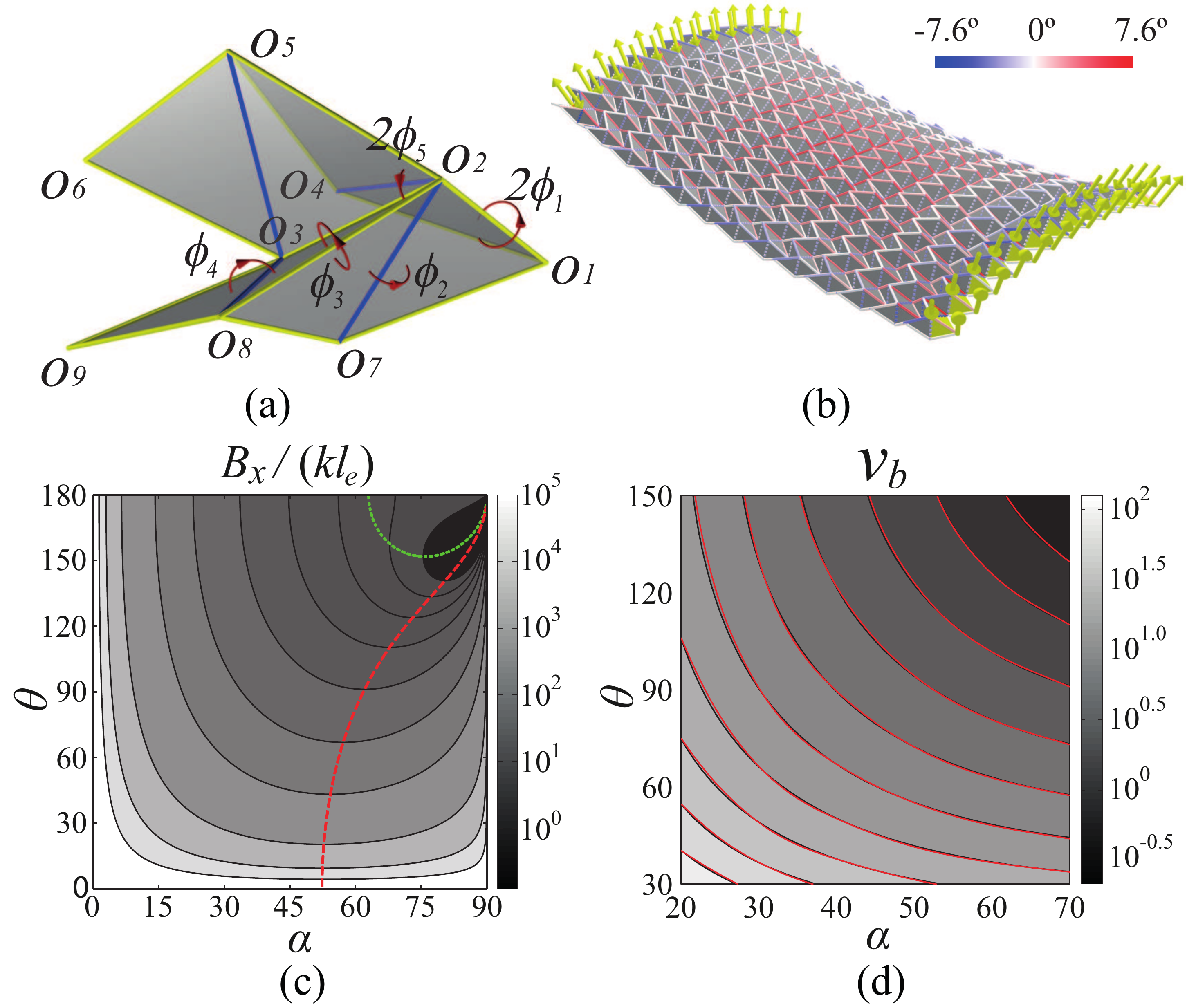}}
\caption{Out of plane bending response of a unit cell. (a) The plaquettes deformations about each fold are symmetric about the plane $O_1O_2O_3$, so that the angles $2\phi_1$, $\phi_2$, $\phi_3$, $\phi_4$ and $2\phi_5$ correspond to rotations about the axes $\overrightarrow{O_1O_2}$, $\overrightarrow{O_7O_2}$, $\overrightarrow{O_2O_8}$, $\overrightarrow{O_8O_3}$ and $\overrightarrow{O_3O_2}$ respectively. (b) Numerical simulation of the bending of a Miura-ori plate with $\alpha = 45^o$ and $\theta = 90^o$. Force dipoles are shown by yellow arrows. Color of the folds indicates the value of deformation angles. (c) Contour plot of dimensionless bending stiffness $B_x/(kl_e)$ corresponding to pure bending of a unit cell. The green dotted curve and red dashed curve indicate the optimal design angle pairs that correspond to the local minima of $B_x|_{\alpha}$ and $B_x|_{\theta}$ respectively. (d) Contour plot of bending Poisson's ratio. The gray scale plot is from the analytic expression \ref{nub} and the red curves are extracted from simulation results. In our simulations, we use a plate made of $21$ by $21$ unit cells and vary $\alpha$ from $20^o$ to $70^o$, $\theta$ from $30^o$ to $150^o$ both every $10^o$.}
\label{Figure3}
\end{figure}

Given the bending behavior of a unit cell, we now turn to a complementary perspective to derive an effective continuum theory for a Miura-ori plate that consists of many unit cells.
Our calculations for the unit cell embodied in (\ref{kappa}) show that $\kappa_x/\kappa_y$ is only a function of the design angles $\alpha$ and $\theta$, and independent of deformation angles, i.e. one cannot independently control $\kappa_x$ and $\kappa_y$. Physically, this means that cylindrical deformations are never feasible, and locally the unit cell is always bent into a saddle. Mathematically, this means that the stiffness matrix of the two-dimensional orthogonal plate \cite{Ventsel2001} is singular, and has rank 1. In the continuum limit, this implies a remarkable result: the Miura plate can be described completely by a 1-dimensional beam theory instead of a 2-dimensional plate theory.

To calculate the bending response of a unit cell, we consider the bending stiffness per unit width of a single cell in the $x$ direction $B_x$. Although the bending energy is physically stored in the $8$ discrete folds, it may also be effectively considered as stored in the entire unit cell that is effectively bent into a sheet with curvature $\kappa_x$. Equating the two expressions allows us to derive $B_x$ ({\bf A}-3.4). In general, $B_x$ depends on multiple deformation angles as they are not necessarily coupled, although here, we only study the ``pure bending" case ({\bf A}-3.5), where a row of unit cells aligned in the $x$ direction undergo the same deformation and the stretching is constrained, i.e. $\phi_1=0$ for all cells, and then $\phi_2=\phi_4$ must be satisfied. In this well-defined case of bending, $B_x$ is solely dependent on the design angles, so that
\begin{equation}
\begin{split}
B_x(\alpha,\theta)=&kl_e \left[2 +16 \sin^3\frac{\alpha }{2} + \left(1-\frac{2 \cos\alpha}{1-\xi ^2}\right)^2\right]\\
&\cot\left(\frac{\theta }{2}\right)\frac{(1-\xi^2)^{3/2}}{2\xi^2\cos\alpha\sin\alpha\cos(\theta/2)},
\end{split}
\end{equation}
{as shown in Fig.\ref{Figure3}c, and we have assumed that all the elastic hinges in a cell have the same stiffness.}

Just as there are optimum design parameters that allow us to extremize the in-plane rigidities, we can also  find the optimal design angle pairs that result in the minima of $B_x$, by setting $\partial_{\theta} B_x|_\alpha=0$ and $\partial_\alpha B_x|_{\theta}=0$. This gives us two curves $\theta_m(\alpha)$ and $\alpha_m(\theta)$ respectively shown in Fig. \ref{Figure3}. The green dotted curve $\theta_m(\alpha)$ starts from $(\alpha,\theta) \approx (63.0^o,180^o)$, and ends at $(\alpha,\theta)=(90^o,180^o)$. It is asymptotically approximated by $2.2851(\alpha-1.0995)\approx(\pi-\theta_m)^2$ when $\alpha\rightarrow 63.0^o$. The red curve $\theta_m(\alpha)$ starts from $(\alpha,\theta) \approx (52.3^o,0^o)$, and ends at $(\alpha,\theta)=(90^o,180^o)$, and is asymptotically approximated by $17.7517(\alpha_m -0.9137) \approx \theta ^2$ when $\theta\rightarrow 0^o$.

{The bending stiffness per unit width of a single cell in the $y$ direction $B_y$ ({\bf{A}}-3.4) is related to $B_x$ via the expression for bending Poisson's ratio $\nu_b^2=B_x/B_y$, where $\nu_b$ is defined in (\ref{nub}).} This immediately implies that optimizing $B_y$ is tantamount to extremizing $B_x$.

The deformation response of a complete Miura-ori plate requires a numerical approach because it is impossible to assemble an entire bent plate by periodically aligning unit cells with identical bending deformations in both the $x$ and $y$ direction ({\bf{A}}-4.1). Our model takes the form of a simple triangle-element based discretization of the sheet, in which each edge is treated as a linear spring with stiffness inversely proportional to its rest length. Each pair of adjacent triangles is assigned an elastic hinge with a bending energy quadratic in its deviation from an initial rest angle that is chosen to reflect the natural shape of the Miur-ori plate. We compute the elastic stretching forces and bending torques in a deformed mesh \cite{Bridson2003, Grinspun2004},  assigning a stretching stiffness that is six orders of magnitude larger than the bending stiffness of the adjacent facets, so that we may deform the mesh nearly isometrically ({\bf{A}}-4.2). When our numerical model of a Miura-ori plate is bent by applied force dipoles along its left-right boundaries, it deforms into a saddle (Fig.\ref{Figure3}b). In this state, asymmetric inhomogeneous twisting arises in most unit cells; indeed this is the reason for the failure of averaging for this problem since different unit cells deform differently. This is in contrast with the in-plane case, where the deformations of the unit cell are affinely related to those of the entire plate.

To compare the predictions for the bending Poisson's ratio $\nu_b$ of the one-dimensional beam theory with those determined using our simulations, in  Fig.\ref{Figure3}d we plot $\nu_b$ from (\ref{nub}) (the gray scale contour plot) based on a unit cell and $\nu_b$ extracted at the center of the bent Miura-ori plate from simulations (the red curves). We see that these two results agree very well, because the unit cell in the center of the plate does have a symmetry plane so that only symmetric bending and in-plane stretching modes are activated, consistent with the assumptions underlying (\ref{nub}). ({\bf{A}}-4.2.)

Our physical analysis of the properties of these folded structures, mechanical metamaterials that might be named { Orikozo}, from the Japanese for Folded Matter are rooted in geometry of the unit cell as characterized by a pair of design angles $\alpha$ and $\theta$ together with its symmetry and the constraint of isometric deformations. It leads to simple expressions for the linearized planar stretching rigidities $K_x$, $K_y$, and non-planar bending rigidities $B_x$ and $B_y$. Furthermore, we find that the in-plane Poisson's ratio $\nu_{_{wl}}<0$, while the out-of-plane bending Poisson ration $\nu_b>0$, an unusual combination that is not seen in simple materials, satisfying the general relation i.e. $\nu_{_{wl}}=-\nu_b$;  a consequence of geometry alone. Our analysis also allows us to pose and solve a series of design problems to find the optimal designs of the unit cell  that lead to extrema of stretching and bending rigidities as well as contraction/expansion ratios of the system. This paves the way for the use of optimally designed Miura-ori patterns in such passive settings as three-dimensional nanostructure fabrication \cite{Barbastathis2006}, and raises the possibility of optimal control of actuated origami-based materials in soft robotics {\cite{Wood2010}} and elsewhere using the simple geometrical mechanics approaches that we have introduced here.

We thank the Wood lab for help with laser cutting to build the paper Miura-ori plates shown in Figure \ref{Figure1}, and the Wyss Institute and the Kavli Institute for support, and Tadashi Tokieda for many discussions and the suggestion that these materials be dubbed { Orikozo}.

\section{Appendix}
\section{1. Geometry and Kinematics}
 \setcounter {equation} {0}
 \setcounter {figure} {0}
 \makeatletter
 \renewcommand{\thefigure}{A.\@arabic\c@figure}
 \renewcommand{\theequation}{A.\@arabic\c@equation}

Before we discuss the coupled deformations of the plate embodied functionally as $\beta(\alpha,\theta)$, we investigate the case when $\alpha=\pi/2$ corresponding to an orthogonally folded map that can only be completely unfolded first in one direction and then another, without bending or stretching the sheet except along the hinges. Indeed, when $\alpha=\pi/2$ and $\theta\neq\pi$, Eq. \eqref{Figure2geo} reduces to $\beta=0$, $l=0$ and $h=l_1$, the singular limit when Miura-ori patterned sheets can not be unfolded with a single diagonal pull. Close to this limiting case, when the folds are almost orthogonal, the Miura-ori pattern can remain almost completely folded in the $x$ direction ($\beta$ changes only by a small amount) while unfolds in the $y$ direction as $\theta$ is varied over a large range, only to expand suddenly in the $x$ direction at the last moment. This observation can be explained by expanding Eq. \eqref{Figure2geo} asymptotically as $\alpha\rightarrow\pi/2$ and $\theta\rightarrow\pi$, which yields $\beta\approx\pi-\epsilon/\delta$, $l\approx l_1(2-(\epsilon/\delta)^2/4)$, $w\approx l_2(2-\delta^2-\epsilon^2/4)$ and $h\approx l_1\epsilon/(2\delta)$, where $\delta=\pi/2-\alpha$ and $\epsilon=\pi-\theta$. Thus, we see that for any fixed small constant $\delta$, only when $\epsilon < \delta$, do we find that $\beta\rightarrow\pi$, $l\rightarrow2l_2$ and $h\rightarrow0$, leading to a sharp transition in the narrow neighborhood  ($\sim\delta$) of $\theta=\pi$ as $\alpha \rightarrow \pi/2$ (Fig.\ref{kinematics}a), consistent with our observations.

\begin{figure}
\centerline{\includegraphics[width=1\textwidth]{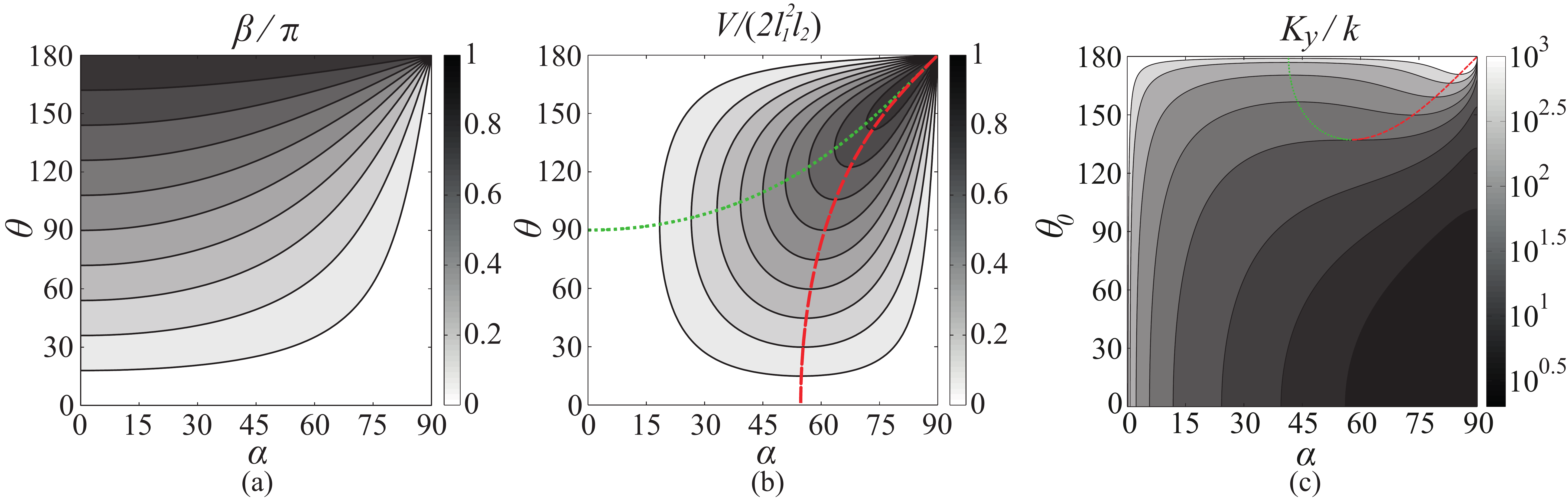}}
\caption{Geometry of the unit cell as a function of $\alpha$ and $\theta$. (a) The folding angle $\beta$ increases as $\theta$ increases and decreases as $\alpha$ increases. The transition becomes sharper as $\alpha \approx \pi/2$, and when $\alpha=\pi/2$, $\beta=0$ independent of $\theta$, i.e. the unfolding (folding) of folded (unfolded) of maps with $N$ orthogonal folds has $2^N$ decoupled possibilities. (b) Effective dimensionless volume $V/(2l_1^2l_2)$. The green dotted curve $\theta_m(\alpha)$ indicates the optimal design angle pairs that correspond to the maximum $V|_\alpha$. The red dashed curve $\alpha_m(\theta)$ indicates the optimal design angle pairs that correspond to the maximum $V|_\theta$. (c) Contour plot of the dimensionless stretching rigidity $K_y/k$. $K_y|_{\alpha}$ is monotonic in $\theta_0$. The green dotted curve indicates the design angle pairs that correspond to the minima of $K_y|_{\theta_0}$. The red dashed curve indicates the design angle pairs that correspond to the maxima of $K_x|_{\theta_0}$. See the text for details.}
\label{kinematics}
\end{figure}

More generally, we start by considering the volumetric packing of Miura-ori characterized by the effective volume of a unit cell
$V\equiv l\times w\times h= 2l^{2}_{1}l_{2} \zeta^2 \sin\theta \sin\alpha \tan\alpha$,
which vanishes when $\theta=0$, $\pi$. To determine the conditions when the volume is at an extremum for a fixed in-plane angle $\alpha$, we set $\partial_{\theta} V|_\alpha=0$ and find that the maximum volume
\begin{equation}
V_{max}|_\alpha =  2l^{2}_{1}l_{2}\sin^2\alpha \ \ \mbox{at}\ \ \theta_m=\cos^{-1}\left(\frac{\cos 2\alpha-1}{\cos{2\alpha}+3}\right),
\end{equation}
shown as a red dashed line in Fig.\ref{kinematics}b. Similarly, for a given dihedral angle $\theta$, we may ask when the volume is extremized as a function of $\alpha$? Using the condition  $\partial_{\alpha} V|_\theta=0$ shows that the maximum volume is given by
\begin{equation}
V_{max}|_\theta =  \frac{4 l_1^2 l_2 \cos\alpha_m \left(\sqrt{5+4 \cos\theta }-3\right) \cot^2\left(\theta/2\right) \sin\theta}{\sqrt{5+4 \cos\theta }-3-2 \cos\theta}
\end{equation}
at
%\[\alpha_m = \cos^{-1}\left(\sqrt{\frac{2+\cos\theta -\sqrt{5+4 \cos\theta }}{\cos\theta-1 }}\right),\]
\[\alpha_m = \cos^{-1}\left[\sqrt{\left(2+\cos\theta -\sqrt{5+4 \cos\theta }\right)/(\cos\theta-1 )}\right],\]
shown as a red dashed line in Fig.\ref{kinematics}b. These relations for the maximum volume as a function of the two angles that characterize the Miura-ori allow us to manipulate the configurations for the lowest density in such applications as packaging for the best protection. In the following sections, we assume each plaquette is a rhombus, i.e. $l_1=l_2=l_e$, to keep the size of the algebraic expressions manageable, although it is a relatively straightforward matter to account for variations from this limit.

\section{2. In-plane stretching response of a Miura-ori plate}
\subsection{2.1 Poisson's ratio related to height changes}
Poisson's ratios related to height changes, $\nu_{_{hl}}$ and $\nu_{_{wl}}$ read
\begin{align}
\begin{split}
&\nu_{_{hl}}=\nu_{_{lh}}^{-1}\equiv-\frac{dh/h}{dl/l}=\cot^2\alpha\sec^2\frac{\theta}{2},\\
&\nu_{_{hw}}=\nu_{_{wh}}^{-1}\equiv-\frac{dh/h}{dw/w}=\zeta^2\tan^2\frac{\theta}{2}.
\end{split}
\label{nu_h}
\end{align}
which are both positive, and monotonically increasing with $\theta$ and $\alpha$. Expansion of $\nu_{_{hl}}$ in Eq. \eqref{nu_h} shows that $\nu_{_{hl}}|_{\theta\rightarrow\pi}\sim (\pi-\theta)^{-2}$ and thus $\nu_{_{hl}}|_{\alpha}\in[\cot^2\alpha,\infty)$, while $\nu_{_{hl}}|_{\alpha\rightarrow0}\sim\alpha^{-2}$ and thus $\nu_{_{hl}}|_{\theta}\in(\infty,0]$. Similarly, expansion of $\nu_{_{hw}}$ in Eq. \eqref{nu_h} shows that $\nu_{_{hw}}|_{\theta\rightarrow\pi}\sim(\pi-\theta)^{-2}$ and thus $\nu_{_{hw}}|_\alpha \in [0,\infty)$, while $\nu_{_{hw}}|_\theta\in[\tan^2(\theta/2),0]$. Finally, it is worth pointing out that $\nu_{_{hw}}$ has a singularity at $(\alpha,\theta) = (\pi/2,\pi)$.

\subsection{2.2 Stretching stiffness $K_x$ and $K_y$}
Here we derive the expressions for stretching stiffness $K_x$ and $K_y$.

The expression for the potential energy of a unit cell deformed by a uniaxial force $f_x$ in the $x$ direction is given by
\begin{equation}
H=U-\int_{\theta_0}^{\theta} f_x \frac{dl}{d \theta'} d\theta',
\label{H}
\end{equation}
where the unit cell length $l$ is defined in Eq. \eqref{Figure2geo}. The elastic energy of a unit cell $U$ is stored only in the elastic hinges which allow the plaquettes to rotate, and is given by
\begin{equation}
U=kl_e(\theta-\theta_0)^2+kl_e(\beta-\beta_0)^2,
\label{U}
\end{equation}
where  $k$ is the hinge spring constant, and $\theta_0$ and $\beta_0$ $(=\beta(\alpha,\theta_0))$ are the natural dihedral angles in the undeformed state. The external force $f_x$ at equilibrium state is obtained using the condition that the first  variation $\delta H/\delta \theta=0$, which reads
\begin{equation}
f_x=\frac{dU/d\theta}{dl/d\theta}=2k\frac{(\theta-\theta_0)+(\beta-\beta_0)\varpi(\alpha,\theta)}{\eta(\alpha,\theta)},
\end{equation}
where $U$ is defined in Eq. \eqref{U}, $l$ is defined in Eq. \eqref{Figure2geo}, and in addition
\begin{equation}
\varpi(\alpha,\theta)=\frac{\cos\alpha}{1-\xi^2} \quad \mbox{and} \quad \eta(\alpha,\theta)=\frac{\cos\alpha\sin^2\alpha\sin\theta}{2(1-\xi^2)^{3/2}}.
\label{notation1}
\end{equation}
The stretching rigidity  associated with the $x$ direction is thus given by
\begin{equation}
K_x(\alpha,\theta_0)\equiv\frac{df_x}{d\theta}\bigg\vert_{ _{\theta_0}}
=4k\frac{(1-\xi_0^2)^{2}+\cos^2\alpha}{(1-\xi_0^2)^\frac{1}{2}\cos\alpha\sin^2\alpha\sin\theta_0},
\label{Kx}
\end{equation}
where $\xi_0 = \xi(\alpha,\theta_0)$.

Similarly, the uniaxial force in the $y$ direction in a unit cell at equilibrium is
\begin{equation}
f_y=\frac{dU/d\theta}{dw/d\theta}=2k\frac{(\theta-\theta_0)+(\beta-\beta_0)\varpi(\alpha,\theta)}{\sin\alpha\cos(\theta/2)},
\end{equation}
where $w$ is defined in Eq. \eqref{Figure2geo} and $\varpi$ is defined in Eq. \eqref{notation1}. The stretching rigidity in $y$ direction is thus given by
\begin{equation}
K_y(\alpha,\theta_0)\equiv\frac{df_y}{d\theta}\bigg\vert_{ _{\theta_0}}=2k\frac{(1-\xi_0^2)^2+\cos^2\alpha}{(1-\xi_0^2)^2\sin\alpha\cos(\theta_0/2)},
\end{equation}
of which the contour plot is show in Fig. \ref{kinematics}c.

\subsection{2.3 Asymptotic cases for optimal design angles}
The expressions in Section 2.2 allow us to derive in detail all the asymptotic cases associated with the optimal pairs of design angles which correspond to the extrema of stretching rigidities $K_x$ and $K_y$. For simplicity, we use $(\alpha,\theta)$ instead of $(\alpha,\theta_0)$ to represent the design angle pairs when the unit cell is at rest.

\begin{enumerate}
\item
Expanding $\partial_\theta K_x$ in the neighborhood of $\alpha=0$ yields
\begin{equation}
\partial_\theta K_x|_{\alpha\rightarrow0}=-\frac{8 \cot\theta  \csc\theta}{\alpha ^2}-\frac{2}{3} \left((3+\cos\theta ) \csc^2\theta \right)+O(\alpha^2).
\label{Kxtheta}
\end{equation}
As $\alpha \rightarrow 0$, $\theta \rightarrow \pi/2$ to prevent a divergence. Continuing to expand Eq. \eqref{Kxtheta} in the neighborhood of $\theta=\pi/2$ and keeping the first two terms yields
\begin{equation}
\partial_\theta K_x|_{\theta\rightarrow\pi/2} =0 \Rightarrow 4(\theta-\pi/2)=\alpha^2.
\label{Kxalpha}
\end{equation}
Therefore in the contour plot of $K_x$ (Fig.3b in the main text), the greed dotted curve is approximated by $4(\theta-\pi/2) = \alpha^2$ in the neighborhood of $\alpha=0$, and is perpendicular to $\alpha=0$ as $\theta$ is quadratic in $\alpha$.

\item
Expanding $\partial_\alpha K_x$ in the neighborhood of $\theta=0$ yields
\begin{equation}
\begin{split}
\partial_\alpha K_x|_{\theta\rightarrow0}=&-\frac{\left[11+20 \cos(2 \alpha)+\cos(4 \alpha )\right] \csc^3\alpha \sec^2\alpha }{2 \theta }-\frac{1}{192} \left\{\left[290+173 \cos(2 \alpha )\right.\right.\\
&\left.\left.+46 \cos(4 \alpha) +3 \cos(6 \alpha )\right] \csc^3\alpha \sec^2\alpha \right\} \theta +O(\theta^2).
\end{split}
\label{Kxalpha}
\end{equation}
The numerator of the leading order in Eq. \eqref{Kxalpha} has to vanish as $\theta\rightarrow0$ to keep the result finite, which results in a unique solution $\alpha^*= \cos^{-1}\left(\sqrt{\sqrt{5}-2}\right)$ in the domain $\alpha\in(0,\pi/2)$. Continuing to expand Eq. \eqref{Kxalpha} in the neighborhood of $\alpha=\alpha^*$ and only keeping the first two terms yields
\begin{equation}
\partial_\alpha K_x =0|_{\alpha\rightarrow\alpha^*} \Rightarrow 4\sqrt{5(1+\sqrt{5})}(\alpha-\alpha^*)=\theta^2.
\end{equation}
so the red dashed curve in the contour plot of $K_x$ (Fig.3b in the main text) is perpendicular to $\theta=0$.

\item
Similarly, Expansion of $\partial_\alpha K_y$ near $\theta=\pi$ yields
\begin{equation}
\begin{split}
\partial_\alpha K_y|_{\theta\rightarrow\pi} =&\frac{[-1+16 \cos(2 \alpha )+\cos(4 \alpha )] \csc^2\alpha \sec^3\alpha}{2 (\theta -\pi )}+\frac{1}{192} [638-737 \cos(2 \alpha )+\\
&162 \cos(4 \alpha )+\cos(6 \alpha )] \csc^2\alpha  \sec^5\alpha (\theta -\pi )+O[(\theta -\pi)^3].
\end{split}
\label{Kyalpha}
\end{equation}
Allowing for a well behaved limit at leading order as $\theta\rightarrow\pi$ requires $-1+16 \cos(2 \alpha )+\cos(4 \alpha )=0$ and yields $\alpha^*=\cos^{-1}\left(\sqrt{ ( \sqrt{17}-3)/2}\right)$ as the unique solution when $\alpha$ is an acute. Again expanding Eq. \eqref{Kyalpha} in the neighborhood of $\theta = \pi$,  and only keeping the first two terms yields
\begin{equation}
\partial_\alpha K_y|_{\alpha\rightarrow\alpha^*}=0\Rightarrow 2 \sqrt{1 + \sqrt{17}}(\alpha_m-\alpha^*) = (\pi-\theta)^2.
\end{equation}
So the green dotted curve in the contour plot of $K_y$ (Fig. \ref{kinematics}c) is approximated by $2 \sqrt{1 + \sqrt{17}}(\alpha_m-\alpha^*) = (\pi-\theta)^2$ near $\alpha=\alpha^*$, and is perpendicular to $\theta=\pi$. The point where the green curve ends satisfies the condition
\begin{equation}
\partial _{\alpha }K_y=0\quad\text{and}\quad \partial _{\alpha }\left(\partial _{\alpha }K_y\right)=0
\end{equation}
and numerical calculation gives us the coordinates of this critical point as
\begin{equation}
\theta = 2.39509,\quad \text{and} \quad \alpha = 1.00626.
\end{equation}
The red dashed curve (Fig. \ref{kinematics}c) starting at this point shows a collection of optimal design angle pairs $(\alpha,\theta)$ where $K_y|_\theta$ is locally maximal.
\end{enumerate}

\section{3. Out-of-plane bending response of a Miura-ori plate}
\subsection{3.1 Minimum model for isometric bending}
Here we show that planar folding is the only geometrically possible motion under the assumption that the unit cell deforms isometrically, i.e. with only rotations of the rhombus plaquettes about the hinges. To enable the out-of-plane bending mode, the minimum model for isometric deformations requires the introduction of $1$ additional diagonal fold into each plaquette, and this follows from the explanation below.

Suppose the plane $O_1O_2O_5O_4$ (Fig.\ref{unitcell_bent}a) is fixed to eliminate all rigid motions, for any dihedral angle $\theta$, the orientation of plane $O_1O_2O_8O_7$ is determined. However, the other two rhombi $O_2O_5O_6O_3$ and $O_2O_3O_9O_8$ are free to rotate about axis $O_2O_5$ and $O_2O_8$ respectively and sweep out two cones which intersect at $O_2O_3$ and $O_2O_3'$. Fig.\ref{unitcell_bent}a shows the two possible configurations of a unit cell determined from the two intersections, the yellow part being the red part that has been flipped about a plane of symmetry. The unit cell in red is the only nontrivial Miura pattern, so that for any given $\theta$, there is a unique configuration of the unit cell corresponding to it. Any continuous change in $\theta$ results in the unit cell being expanded or folded but remaining planar, in which case, $O_1, O_4, O_7, O_3, O_6$ and $O_9$ also remain coplanar. In order to enable the bending mode of the unit cell, the planarity of each plaquette must be violated. In the limit where the plaquette thickness $t\ll1$  the stretching rigidity ($\sim t$) is much larger than the bending rigidity ($\sim t^3$), with $t$ being the thickness of a plaquette, while the energy required to bend a strip of ridge is $5$ times of that required to stretch it according to the asymptotic analysis of the $F\ddot{o}ppl-von\ K\acute{a}rm\acute{a}n$ equations \cite{Lobkovsky1996}. Therefore, the rigid ridge/fold is an excellent approximation for out-of-plane bending when $t\ll1$. Then, to get a bent shape in a unit cell and thence in a Miura-ori plate, we must introduce an additional fold into each rhombus to divide it into two elastically hinged triangles (Fig.\ref{unitcell_bent}b). As a result, 4 additional degrees of freedom are introduced in each unit cell. The deformed state can either be symmetrical about the plane $O_1O_2O_3$ (Fig.\ref{unitcell_bent}c) corresponding to a bending mode, or unsymmetrical corresponding to a twisting mode. Here, we are only interested in the bent state, in which the rotation angle $\phi_2$ about the axis $\overrightarrow{O_2O_4}$, and $\phi_4$ about the axis $\overrightarrow{O_3O_5}$, are the same as the rotations about $\overrightarrow{O_7O_2}$ and $\overrightarrow{O_8O_3}$ respectively. The rotation angles about the axis $\overrightarrow{O_1O_2}$, $\overrightarrow{O_3O_2}$ and $\overrightarrow{O_2O_5}$ are $2\phi_1$, $2\phi_5$ and $\phi_3$ respectively. ($\overrightarrow{}$ indicates the direction.)

\begin{figure}
\centerline{\includegraphics[width=1\textwidth]{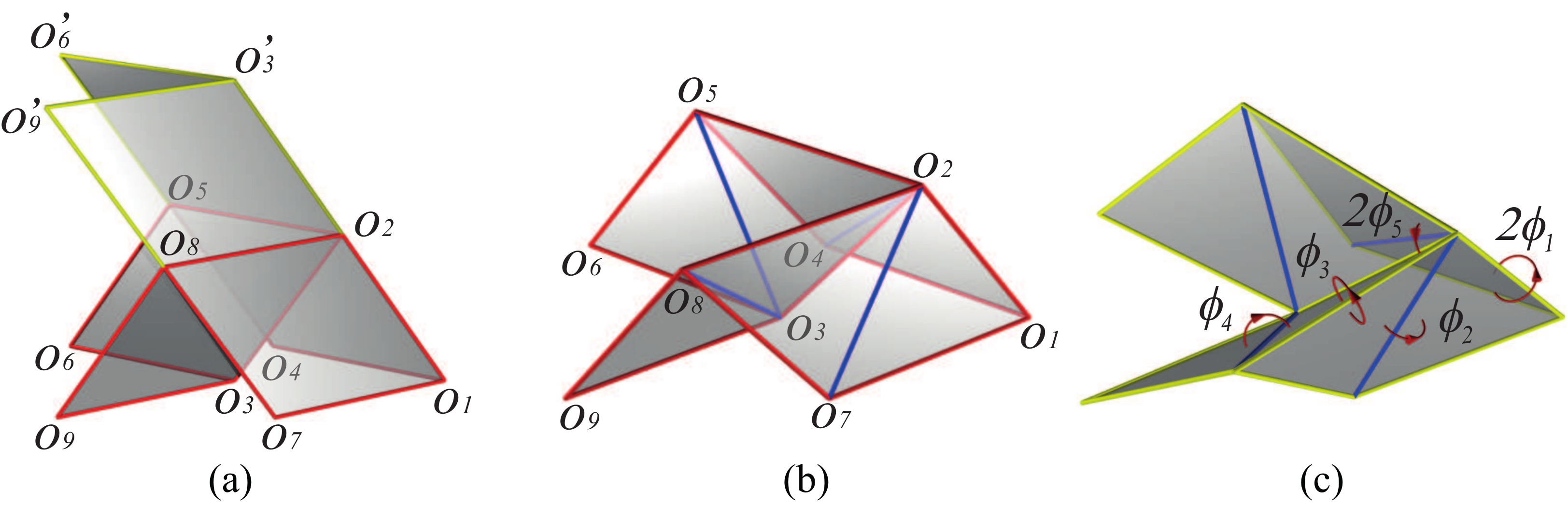}}
\caption{Bending of a unit cell. (a) The two configurations of a unit cell for any given $\theta$ if each plaquette is a rigid rhombus. The only possible motion is in-plane stretching. The yellow plaquettes illustrate the trivial configuration of two rigid plaquettes and the red ones show the typical configuration of a Miura-ori unit cell. (b) The undeformed state. An additional fold along the short diagonal is introduced to divide each rhombus into 2 elastically hinged triangles. (c) Symmetrically bent state. The bending angles around axis $\overrightarrow{O_2O_4}$ and $\overrightarrow{O_3O_5}$ are the same as those around $\overrightarrow{O_7O_2}$ and $\overrightarrow{O_8O_3}$ respectively.}
\label{unitcell_bent}
\end{figure}

\subsection{3.2 Curvatures and the bending Poisson's ratio when short folds are introduced}
Here we derive expressions for the coordinates of every vertex of the unit cell after bending in the linear deformation regime, from which curvatures in the two principle directions $\kappa_x$, $\kappa_y$ and the bending Poisson's ratio $\nu_b = -\kappa_y/\kappa_x$ can be calculated.

To do so, we first need to know the transformation matrix associated with rotation about an arbitrary axis. The rotation axis is defined by a point $\{a,b,c\}$ that it goes through and a direction vector $<u,v,w>$, where $u$, $v$, $w$ are directional cosines. Suppose a point $\{x_0,y_0,z_0\}$ rotates about this axis by an infinitesimal small angle $\omega$ ($\omega \ll1$), and reaches the new position $\{x,y,z\}$. Keeping only the leading order terms of the transformation matrix, we find that the new position $\{x,y,z\}$ is given by
\begin{equation}
\begin{split}
&x = x_0 + (-c v + b w - w y_0 + v z_0)\omega,\\
&y = y_0 + (c u - a w + w x_0 - u z_0)\omega,\\
&z = z_0 + (-b u + a v - v x_0 + u y_0)\omega.
\end{split}
\label{transform}
\end{equation}

Given Eq. \eqref{transform}, we are ready to calculate the coordinates of all vertices in the bent sate. Assuming that the origin is at $O_2$, in the undeformed unit cell, edge $O_1O_2$ is fixed in $xoz$ plane to eliminate rigid motions. Each fold deforms linearly by angle $2\phi_1$, $\phi_2$, $\phi_3$, $\phi_4$ and $2\phi_5$ (see Fig. \ref{unitcell_bent}c) around corresponding axes respectively.  The coordinates of $O_1$ and $O_2$ are
\begin{equation}
\begin{split}
&O_{1x}=\frac{\cos\alpha}{\sqrt{1-\xi^2}},\quad O_{1y}=0,\quad O_{1z}=-\frac{\sin\alpha\cos(\theta /2)}{\sqrt{1-\xi ^2}};\\
&O_{2x}=0,\quad O_{2y}=0,\quad O_{2z}=0.
\end{split}
\label{O1O2}
\end{equation}

The coordinates of $O_3$ after bending are
\begin{equation}
\begin{split}
O_{3x} =&-\frac{\cos\alpha}{\sqrt{1-\sin^2\alpha \sin^2(\theta/2)}}-\frac{\cos(\alpha/2)\sin\alpha \sin\theta }{\sqrt{3-\cos(2 \alpha ) (\cos\theta-1)+\cos\theta }}\phi_2+\frac{\sin^2\alpha \sin\theta }{2 \sqrt{1-\sin^2\alpha \sin^2(\theta /2)}}\phi_3,\\
O_{3y} = &-\frac{4\cos(\theta/2)\sin(2\alpha)}{3+\cos(2\alpha)+2\cos\theta\sin^2\alpha}\phi_1+\frac{\csc(\theta/2)[-\sin\alpha+\sin(2\alpha)+\sin^3\alpha\sin^2(\theta/2)]
\sin\theta}{[3+\cos(2\alpha)+2\cos(\theta)\sin^2\alpha]\sin(\alpha/2)}\phi_2\\
&+\cos(\theta/2)\sin(\alpha)\phi_3,\\
O_{3z} =&-\frac{\cos(\theta/2)\sin\alpha}{\sqrt{1-\sin^2\alpha \sin^2(\theta/2)}}+\frac{2\cos\alpha\cos(\alpha/2) \sin(\theta/2) }{\sqrt{3-\cos(2 \alpha ) (\cos\theta-1)+\cos\theta }}\phi_2-\frac{\cos\alpha \sin\alpha\sin(\theta/2) }{\sqrt{1-\sin^2\alpha \sin^2(\theta /2)}}\phi_3.
\end{split}
\end{equation}

The coordinates of $O_4$ after bending are
\begin{equation}
\begin{split}
O_{4x} = &\frac{\cos\alpha+\sin^2\alpha \sin^2(\theta/2)-1}{\sqrt{1-\sin^2\alpha \sin^2(\theta/2)}}-\frac{\sin^2\alpha\sin\theta }{\sqrt{3-\cos(2 \alpha ) (\cos\theta-1)+\cos\theta }}\phi_1,\\
O_{4y} = &\sin\alpha\sin(\theta/2)-\cos(\theta/2)\sin\alpha\phi_1,\\
O_{4z} = &-\frac{\cos(\theta/2)\sin\alpha}{\sqrt{1-\sin^2\alpha \sin^2(\theta/2)}}-\frac{2\cos\alpha\sin\alpha \sin(\theta/2) }{\sqrt{3-\cos(2 \alpha ) (\cos\theta-1)+\cos\theta }}\phi_1.
\end{split}
\label{O4}
\end{equation}

The coordinates of $O_5$ after bending are
\begin{equation}
\begin{split}
O_{5x} = &-\sqrt{1-\sin^2\alpha \sin^2(\theta /2)}-\frac{\sin^2\alpha \sin\theta}{\sqrt{3-\cos(2 \alpha ) (\cos\theta -1)+\cos\theta}}\phi_1\\
&+\frac{\sin^2\alpha \sin\theta}{2 \sqrt{3-\cos(2 \alpha ) (\cos\theta -1)+\cos\theta } \sin(\alpha/2)}\phi_2,\\
O_{5y} = & \sin\alpha \sin(\theta/2)-\cos(\theta/2) \sin\alpha \phi_1+\frac{\cos(\theta/2) \sin\alpha }{2 \sin(\alpha /2)}\phi_2,\\
O_{5z} = &-\frac{2 \cos\alpha \sin\alpha \sin(\theta/2)}{\sqrt{3-\cos(2 \alpha) (\cos\theta-1)+\cos\theta }}\phi_1+\frac{\cos\alpha \sin\alpha \sin(\theta/2)}{\sqrt{3-\cos(2 \alpha) (\cos\theta-1)+\cos\theta } \sin(\alpha /2)}\phi_2.
\end{split}
\label{O5}
\end{equation}

The coordinates of $O_6$ after bending are
\begin{equation}
\begin{split}
O_{6x}= &\frac{\sin^2\alpha \sin^2(\theta/2)-\cos\alpha-1}{\sqrt{1-\sin^2\alpha \sin^2(\theta/2)}}-\frac{\sin^2\alpha\sin\theta }{\sqrt{3-\cos(2 \alpha ) (\cos\theta-1)+\cos\theta }}\phi_1\\
&+\frac{\sin^2\alpha \sin\theta }{2 \sqrt{1-\sin^2\alpha \sin^2(\theta /2)}}\phi_3-\frac{\sin\alpha\sin\theta\cos(\alpha/2) }{\sqrt{3-\cos(2 \alpha ) (\cos\theta-1)+\cos\theta }}\phi_4,\\
O_{6y} = &\sin\alpha \sin(\theta /2)+\frac{4 \cos(\theta /2) \sin\alpha [\sin^2\alpha \sin^2(\theta /2)-1-2 \cos\alpha ]}{3+\cos(2 \alpha) +2 \cos\theta  \sin^2\alpha}\phi_1 \\
&+ \frac{8 \cos\alpha \cos(\theta /2) \cos(\alpha /2)}{3+\cos(2 \alpha )+2 \cos\theta  \sin^2\alpha}\phi_2+\cos(\theta /2)\sin\alpha \phi_3 -\cos(\theta /2) \cos(\alpha/2)\phi_4,\\
O_{6z} = &-\frac{\cos(\theta /2) \sin\alpha }{\sqrt{1-\sin^2\alpha \sin^2(\theta /2)}}-\frac{2 \cos\alpha \sin\alpha \sin(\theta /2)}{\sqrt{3-\cos(2 \alpha ) (\cos\theta -1)+\cos\theta}}\phi_1\\
&+\frac{\csc(\alpha/2) \sin(2 \alpha ) \sin(\theta/2)}{\sqrt{3-\cos(2 \alpha) (\cos\theta -1)+\cos\theta }}\phi_2-\frac{\cos\alpha  \sin\alpha \sin(\theta/2)}{\sqrt{1-\sin^2\alpha \sin^2(\theta/2)}}\phi_3\\
&+\frac{2\cos\alpha\cos(\alpha /2) \sin(\theta /2)}{\sqrt{3-\cos(2 \alpha) (\cos\theta -1)+\cos\theta } }\phi_4.
\end{split}
\label{O6}
\end{equation}

The coordinates of $O_7$, $O_8$ and $O_9$ after bending are
\begin{equation}
\begin{split}
&\{O_{7x},O_{7y},O_{7z}\}=\{O_{4x},-O_{4y},O_{4z}\},\quad \{O_{8x},O_{8y},O_{8z}\}=\{O_{5x},-O_{5y},O_{5z}\},\\ &\{O_{9x},O_{9y},O_{9z}\}=\{O_{6x},-O_{6y},O_{6z}\}.
\end{split}
\end{equation}

Due to symmetry, $O_3$ must lie in the $xoz$ plane after bending, so $O_{3y}=0$, from which $\phi_3$ and $\phi_5$ can be expressed as a function of $\phi_1$ and $\phi_2$,
\begin{align}
\begin{split}
\phi_3&= \frac{8 \cos\alpha }{3+\cos(2 \alpha)+2 cos\theta \sin^2\alpha}\phi_1+\frac{1}{2} \csc\left(\frac{\alpha }{2}\right) \left(1-\frac{8 \cos\alpha}{3+\cos(2 \alpha)+2 \cos\theta \sin^2\alpha}\right)\phi_2.\\
\phi_5&= \phi_1 - \frac{1}{2}\csc\left({\frac{\alpha}{2}}\right)\phi_2
\end{split}
\label{phis}
\end{align}

The curvature of the unit cell in the $x$ direction is defined as the dihedral angle formed by rotating plane $O_4O_1O_7$ to plane $O_6O_3O_9$ projected onto the $x$ direction over the unit length $l$. The sign of the angle follows the right-hand rule about the $y$ axis. The dihedral angle between plane $O_4O_1O_7$ and plane $xoy$ is
\begin{equation}
\Omega_{417} = \frac{O_{4z}-O_{1z}}{\sqrt{1-\xi^2}} = -\frac{4 \cos\alpha \sin\alpha \sin(\theta/2)}{3+\cos(2 \alpha)+2 \cos\theta \sin^2\alpha}\phi_1,
\end{equation}
and the dihedral angle between plane $O_3O_6O_9$ and plane $xoy$ is
\begin{equation}
\Omega_{639} = \frac{O_{6z}-O_{3z}}{\sqrt{1-\xi^2}}=\frac{2 [\cos(\alpha/2)+\cos(3 \alpha /2)] [\phi_2+\phi_4-2 \phi_1 \sin(\alpha/2)] \sin(\theta/2)}{3+\cos(2 \alpha) +2 \cos\theta \sin^2\alpha}.
\end{equation}
The curvature $\kappa_x$ hence is
\begin{equation}
\kappa_x = \frac{\Omega_{639}-\Omega_{417}}{l}=\frac{(\phi_2+\phi_4) \cos(\alpha/2)\sin(\theta/2)}{2\sqrt{1-\xi^2}}.
\label{kappax}
\end{equation}

The curvature of the unit cell in the $y$ direction is defined as the dihedral angle between plane $O_4O_5O_6$ and $O_7O_8O_9$ projected onto the $y$ direction over the unit cell width $w$, which is expressed as
\begin{equation}
\kappa_y = -\frac{2O_{5y}-O_{4y}-O_{3y}}{hw} = -\frac{1}{4} (\phi_2+\phi_4) \csc\left(\frac{\alpha }{2}\right) \csc\alpha \csc\left(\frac{\theta }{2}\right) \sqrt{1-\xi^2}.
\label{kappay}
\end{equation}

From Eq. \eqref{kappax} and Eq. \eqref{kappay}, we can calculate the bending Poisson ratio, which is simplified to
\begin{equation}
\nu_b= -\frac{\kappa_y}{\kappa_x}=-1+\csc^2\alpha \csc^2\left(\frac{\theta }{2}\right).
\end{equation}

\subsection{3.3 Curvatures and the bending Poisson's ratio when long folds are introduced}
In Fig.\ref{unitcell_bent}, if we introduce the additional fold along the long diagonal, e.g. $O_1O_5$, instead of the short one, the unit cell can be bent too. In this case, $\phi_2$ and $\phi_4$ are bending angles around axis $\overrightarrow{O_1O_5}$ and $\overrightarrow{O_2O_6}$ respectively. $O_1$, $O_2$ do not change as they are fixed, and coordinates of $O_3$ after bending are
\begin{equation}
\begin{split}
O_{3x} =& -\frac{\cos\alpha}{\sqrt{1-\sin^2\alpha\sin^2(\theta/2)}}+\frac{\sin^2\alpha\sin\theta}{2\sqrt{1-\sin^2\alpha\sin^2(\theta/2)}}\phi_3
-\frac{\sin\alpha\sin(\alpha/2)\sin\theta}{\sqrt{3-\cos(2 \alpha ) (\cos\theta-1)+\cos\theta }}\phi_4,\\
O_{3y} =&-\frac{4\cos(\theta/2)\sin(2\alpha)}{3+\cos(2\alpha)+2\cos\theta\sin^2\alpha}\phi_1+\cos\left(\frac{\theta}{2}\right)\sin(\alpha)\phi_3-
\cos\left(\frac{\theta}{2}\right)\sin\left(\frac{\alpha}{2}\right)\phi_4,\\
O_{3z} =& -\frac{\cos(\theta/2)\sin(\alpha)}{\sqrt{1-\sin^2\alpha\sin^2(\theta/2)}}-\frac{\cos\alpha\sin\alpha\sin(\theta/2)}{\sqrt{1-\sin^2\alpha\sin^2(\theta/2)}}\phi_3
+\frac{2\cos\alpha\sin(\alpha/2)\sin(\theta/2)}{\sqrt{3-\cos(2 \alpha ) (\cos\theta-1)+\cos\theta }}\phi_4.
\end{split}
\end{equation}

The coordinates of $O_4$ after bending are
\begin{equation}
\begin{split}
O_{4x}=&\frac{\cos\alpha-1+\sin^2\alpha \sin^2(\theta/2)}{\sqrt{1-\sin^2\alpha \sin^2(\theta/2)}}-\frac{\sin^2\alpha\sin\theta}{\sqrt{3-\cos(2 \alpha ) (\cos\theta-1)+\cos\theta }}\left[\phi_1-\frac{1}{2}\sec\left(\frac{\alpha}{2}\right)\phi_2\right],\\
O_{4y}=&\sin\alpha\sin(\theta/2)-\cos(\theta/2)\sin\alpha\phi_1+\cos(\theta/2)\sin(\alpha/2)\phi_2,\\
O_{4z} =& -\frac{\cos(\theta/2)\sin\alpha}{\sqrt{1-\sin^2\alpha\sin^2(\theta/2)}}-\frac{2\cos\alpha\sin(\theta/2)\sin(\alpha/2)}{\sqrt{3-\cos(2 \alpha ) (\cos\theta-1)+\cos\theta }}\left[2\cos\left(\frac{\alpha}{2}\right)\phi_1-\phi_2\right].
\end{split}
\end{equation}

The coordinates of $O_5$ after bending are
\begin{equation}
\begin{split}
O_{5x}=&-\sqrt{1-\sin^2\alpha\sin^2\left(\frac{\theta}{2}\right)}-\frac{\sin^2\alpha\sin\theta}{\sqrt{3-\cos(2 \alpha ) (\cos\theta-1)+\cos\theta }}\phi_1,\\
O_{5y}=&\sin\alpha\sin(\theta/2)-\cos(\theta/2)\sin\alpha\phi_1,\\
O_{5z} =&-\frac{\sin(2\alpha)\sin(\theta/2)}{\sqrt{3-\cos(2 \alpha ) (\cos\theta-1)+\cos\theta }}\phi_1.
\end{split}
\end{equation}

The coordinates of $O_6$ after bending are
\begin{equation}
\begin{split}
O_{6x}=&\frac{\sin^2\alpha \sin^2(\theta/2)-\cos\alpha-1}{\sqrt{1-\sin^2\alpha \sin^2(\theta/2)}}-\frac{\sin^2\alpha\sin\theta}{\sqrt{3-\cos(2 \alpha ) (\cos\theta-1)+\cos\theta }}\phi_1+\frac{\sin^2\alpha\sin\theta}{2\sqrt{1-\sin^2\alpha\sin^2(\theta/2)}}\phi_3,\\
O_{6y}=&\sin\alpha \sin\left(\frac{\theta }{2}\right)+\frac{4 \cos(\theta/2) \sin\alpha \left[-1-2 \cos\alpha+\sin^2\alpha \sin^2(\theta/2)\right]}{3+\cos(2 \alpha)+2 \cos\theta \sin^2\alpha}\phi_1+\cos\left(\frac{\theta }{2}\right) \sin\alpha\phi_3,\\
O_{6z} =& -\frac{\cos(\theta/2)\sin\alpha}{\sqrt{1-\sin^2\alpha\sin^2(\theta/2)}}-\frac{\sin(2\alpha)\sin(\theta/2)}{\sqrt{3-\cos(2 \alpha ) (\cos\theta-1)+\cos\theta }}\phi_1-\frac{\cos\alpha\sin\alpha\sin(\theta/2)}{\sqrt{1-\sin^2\alpha\sin^2(\theta/2)}}\phi_3.
\end{split}
\end{equation}

Using the same idea for the long fold case as we did for the short fold, we can also calculate the curvatures in the two principal directions and find that
\begin{equation}
\kappa_x = \frac{\Omega_{639}-\Omega_{417}}{l}=\frac{2 \left[\sin(\alpha/2)-\sin(3\alpha/2)\right]\sin(\theta/2)}{[3+\cos(2\alpha)+2\cos\theta\sin^2\alpha]l}(\phi_2+\phi_4)=\frac{\sin\left(\alpha/2\right) \sin\left(\theta/2 \right)}{2l_e\sqrt{1-\xi^2}}(\phi_2+\phi_4),
\end{equation}
while
\begin{equation}
\kappa_y = -\frac{2O_{5y}-O_{4y}-O_{3y}}{hw} = -\frac{\sqrt{1-\sin^2\alpha \sin^2(\theta/2)}}{2\cos(\alpha/2)w}(\phi_2+\phi_4)=-\frac{\sqrt{1-\xi^2}}{4l_e\cos\left(\alpha/2\right)\xi}(\phi_2+\phi_4).
\end{equation}

Therefore the bending Poisson ratio is
\begin{equation}
\nu_b= -\frac{\kappa_y}{\kappa_x}=-1+\csc^2\alpha \csc^2\left(\frac{\theta }{2}\right),
\end{equation}
which is the same as that of the case when the short folds are introduced.

\subsection{3.4 Bending stiffness $B_x$ and $B_y$}
We are now in a position to derive expressions for the bending stiffness $B_x$ and $B_y$. On one hand, the bending energy is physically stored in the $8$ discrete folds, which can be expressed as $1/2k l_e[4\phi_1^2+4\sin(\alpha/2)\phi_2^2+2\phi_3^2+4\sin(\alpha/2)\phi_4^2+4\phi_5^2]$. On the other hand  from a continuum point of view, the energy may also be effectively considered as stored in the entire unit cell that is bent into the curvature $\kappa_x$, which can be expressed as $1/2B_x w l \kappa_x^2$. Equating the two expressions for the same energy, we can write $B_x$ as
\begin{equation}
B_x = k l_e\frac{4\phi_1^2+4\sin(\frac{\alpha}{2})\phi_2^2+2\phi_3^2+4\sin(\frac{\alpha}{2})\phi_4^2+4\phi_5^2}{w l\kappa_x^2}.
\label{Bx}
\end{equation}

Similarly, the bending stiffness per unit width of a single cell in the $y$ direction is
\begin{equation}
B_y(\alpha,\theta) = k l_e\frac{4\phi_1^2+4\sin(\frac{\alpha}{2})\phi_2^2+2\phi_3^2+4\sin(\frac{\alpha}{2})\phi_4^2+4\phi_5^2}{w l\kappa_y^2}.
\label{By}
\end{equation}

\subsection{3.5 Pure bending}
Finally, we explain the ``pure bending" situation in the main text,  borrowing ideas from notions of the pure bending of a beam where curvature is constant. If we demand that a row of unit cells aligned in the $x$ direction (e.g. the cell $C_1$ and $C_2$ in Fig.\ref{assemble}) undergo exactly the same deformation, this results in $\phi_2=\phi_4$. Furthermore, in this limit, the stretching mode is constrained, so that $\phi_1=0$ for all cells. For this well defined bending deformation, the bending stiffness depends only on the design angles, not on the deformation angles as shown in Eq. \eqref{Bx} and Eq. \eqref{By}.

\section{4. Numerical simulations of the bending response of a Miura-ori plate}

\subsection{4.1 Homogeneous deformation in bent plate is impossible}

Here we explain why it is impossible to assemble an entire bent plate by periodically aligning unit cells with identical bending deformation in both the $x$ and $y$ direction.

In Fig.\ref{assemble}, the 4 unit cells $C_1$, $C_2$, $C_3$ and $C_4$ have identical bending deformations: $C_1$ and $C_2$ align perfectly in the $x$ direction, which requires that $\angle O_4O_1O_7= \angle O_6O_3O_9=\angle O_{11}O_{13}O_{15}$. $C_1$ and $C_3$, $C_2$ and $C_4$ align perfectly in the $y$ direction respectively, which is automatically satisfied by the symmetry of the unit cell. Now the question becomes whether the unit cell $C_3$ and $C_4$ can align together? The answer is no. The reasoning is as follows. $O_3$ and $O_3^{'}$ are symmetric about plane $O_6O_{12}O_{13}$, while $O_3$ and $O_3^{''}$ are symmetric about plane $O_4O_5O_6$. However plane $O_6O_{12}O_{13}$ and plane $O_4O_5O_6$ are not coplanar unless all the deformation angles about the internal folds are zero, which is violated by bending. $O_3^{'}$ and $O_3^{''}$ thus do not coincide. In fact $O_3^{'}=O_3^{''}$ if and only if $O_{3y}=O_{5y}=O_{6y}$, which requires $\phi_2=\phi_4=0$ from Eq. \eqref{O4}, Eq. \eqref{O5}, Eq. \eqref{O6} and Eq. \eqref{phis}. This is the in-plane stretching mode instead of the bending mode. In conclusion, in the bent Miura-ori plate, the deformation must be inhomogeneous.

\begin{figure}[h*]
\centerline{\includegraphics[width=.7\textwidth]{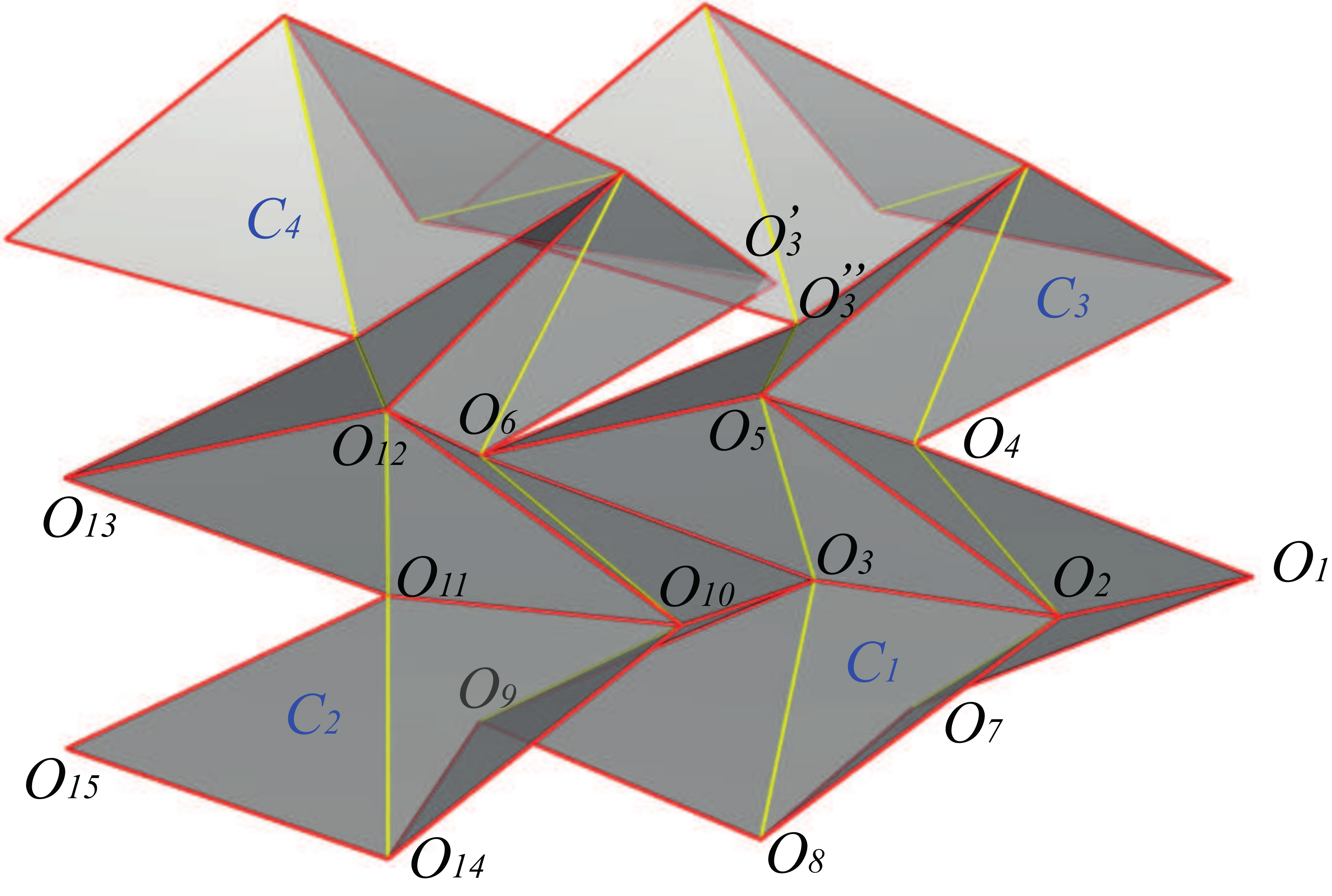}}
\caption{4 unit cells with identical bending deformation cannot be aligned together to form an entire plate. See the text for details.}
\label{assemble}
\end{figure}

\subsection{4.2 Simulation model}
In this subsection, we explain the bending model and the strategies used to bend the Miura-ori plate.
\begin{figure}[h*]
\centerline{\includegraphics[width=1\textwidth]{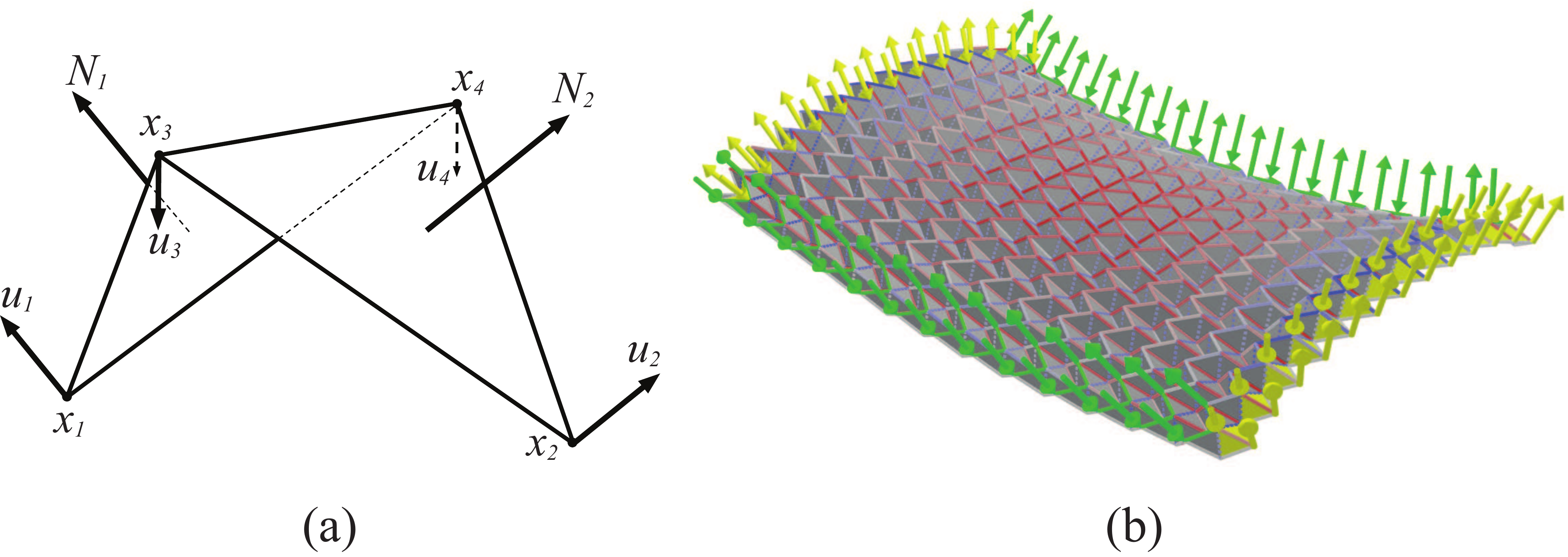}}
\caption{Simulation model. (a) A single bending adjacency. The vectors $u_i$ illustrate the purely geometric bending mode and $N_1$ and $N_2$ are the weighted normals of the adjacent triangles. (b) The left-right bending strategy is shown in yellow and the up-down bending strategy is shown in green. Each arrow represents a force applied to its incident vertex. Left-right force directions bisect the yellow adjacencies and are perpendicular to the shared edge and up-down force directions are normal to the plane spanned by each pair of green edges.}
\label{sim_model}
\end{figure}

We endow these triangulated meshes with elastic stretching and bending modes to capture the in-plane and out-of-plane deformation of thin sheets. The stretching mode simply treats each edge in the mesh as a linear spring, all edges having the same stretching stiffness. Accordingly, the magnitude of the restorative elastic forces applied to each node in a deformed edge with rest length $x_0$ and stretching stiffness $k$ is given by $\frac{k_s}{x_0}(x' - x_0)$ and the energy contained in a deformed edge is given by
\begin{equation}
\frac{k_s}{2x_0}(x' - x_0)^2.
\end{equation}
The $x_0$ term in denominator of the stretching mode ensures mesh-independence. The bending mode is characterized in terms of four vectors $u_1$, $u_2$, $u_3$ and $u_4$, each of which is applied to a node in a pair of adjacent triangles. Defining the weighted normal vectors $N_{1} = (x_1-x_3)\times (x_1-x_4)$ and $N_{2} = (x_2-x_4)\times (x_2-x_3)$ and the shared edge $E = x_4-x_3$, we may write
\begin{equation}
u_1 = |E|\frac{N_1}{|N_1|^2}
\end{equation}
\begin{equation}
u_2 = |E|\frac{N_2}{|N_2|^2}
\end{equation}
\begin{equation}
u_3 = \frac{(x_1-x_4)\cdot E}{|E|}\frac{N_1}{|N_1|^2} + \frac{(x_2-x_4)\cdot E}{|E|}\frac{N_2}{|N_2|^2}
\end{equation}
\begin{equation}
u_4 = -\frac{(x_1-x_3)\cdot E}{|E|}\frac{N_1}{|N_1|^2} - \frac{(x_2-x_2)\cdot E}{|E|}\frac{N_2}{|N_2|^2}.
\end{equation}

The relative magnitudes of these vectors constitute a pure geometric bending mode for a pair of adjacent triangles. For pairs of adjacent triangles that \emph{do not} straddle the fold line, the force on each vertex is given by
\begin{equation}
F_{i} = k_b(\frac{\theta}{2} - \frac{\theta_0}{2})u_{i},
\end{equation}
where $k_b$ is the bending stiffness and $\theta$ is the angle between $N_1$ and $N_2$ that makes each $u_i$ a restorative force. For pairs of adjacent triangles that straddle folds, $\theta_0$ is non-zero and shifts the rest angle of the adjacency to a non-planar configuration. The bending energy contained in a pair of adjacent triangles is given by
\begin{equation}
E_b = k_b\int_{\theta_0}^{\theta} \frac{\theta}{2} - \frac{\theta_0}{2} \mathrm{d}\theta,
\end{equation}
with a precise form of
\begin{equation}
E_b = k_b(\frac{\theta}{2} - \frac{\theta_0}{2})^2,
\end{equation}
which is quadratic in $\theta$ for $\theta\sim  \theta_0$.

We introduce viscous damping so that the simulation eventually comes to rest. Damping forces are computed at every vertex with different coefficients for each oscillatory mode, bending and stretching. We distinguish between these two modes by projecting the velocities of the vertices in an adjacency onto the bending mode, and the velocities of the vertices in an edge onto the stretching mode.

We use the Velocity Verlet numerical integration method to update the positions and velocities of the vertices based on the forces from the bending and stretching model and the external forces from our bending strategies. At any time $t + \Delta t$ during the simulation we can approximate the position $x(t + \Delta t)$ and the velocity $\dot{x}(t + \Delta t)$ of a vertex as
\begin{equation}
\begin{split}
x(t + \Delta t) &= x(t) + \dot{x}(t)\, \Delta t + \frac{1}{2} \ddot{x}(t)\, \Delta t^2,\\
\dot{x}(t + \Delta t) &= \dot{x}(t) + \frac{\ddot{x}(t) + \ddot{x}(t + \Delta t)}{2}\, \Delta t.
\end{split}
\end{equation}
A single position, velocity and accleration update follows a simple algorithm.
\begin{itemize}
\item Compute $x(t + \Delta t)$
\item Compute $\ddot{x}(t + \Delta t)$ using $x(t + \Delta t)$ for stretching and bending forces and $\dot{x}(t)$ for damping forces
\item Compute $\dot{x}(t + \Delta t)$
\end{itemize}
Note that this algorithm staggers the effects of damping on the simulation by $\Delta t$.

\begin{figure}
\includemovie[
  poster,
  toolbar, %same as `controls'
  %label=threeD,
  %text=(3D geometry of a bent Miura plate made of $21$ by $21$ unit cells.),
  3Daac=90.000000, 3Droll=0.000000, 3Dc2c=0.000000 10.000000 0.000000, 3Droo=25.000000, 3Dcoo=-12.104091 0.341133 9.093266,
  3Dlights=CAD,
]{\textwidth}{\textwidth}{Short_Fold_21x21_Alpha60_Theta60_Result_UD.u3d}
\caption{3D geometry of a bent Miura plate made of $21$ by $21$ unit cells with $\alpha=\theta=\pi/3$. For better display purpose, we use an example with pronounced deformation. However in the simulation we have done, we make sure that the radius of curvature is at least 10 times larger than the plate size, such that the deformation is within linear regime. Readers may want to play with different toolbar options to better visualize the geometry. }
\label{3Dgeo}
\end{figure}

In simulation, the Miura-ori plate is made of $21$ by $21$ unit cells, $21$ being the number of unit cells in one direction. $\alpha$ varies from $20^o$ to $70^o$, and $\theta$ varies from $30^o$ to $150^o$, both every $10^o$. We design two bending strategies, each of which corresponds to a pair of opposite boundaries. The left-right bending strategy identifies the adjacencies with $O_2O_3$ shared edges on left boundary unit cells and $O_1O_2$ shared edges on right boundary unit cells (highlighted in yellow in Fig.\ref{sim_model}b). For each of these adjcencies we apply equal and opposite forces to the vertices on their shared edge, the directions of which are determined to lie in the bisecting plane of $O_1O_2O_4$ and $O_1O_2O_7$ (left boundary unit cells) and $O_2O_3O_5$ and $O_2O_3O_8$ (right boundary unit cells) and perpendicular to the shared edge. The up-down bending strategy identifies the top edges of each unit cell on the up and down boundaries of the pattern (shown in green in Fig.\ref{sim_model}b). Each unit cell has one such pair of edges and we apply equal and opposite forces to the not-shared vertices in this pair, the directions of which are normal to the plane spanned by the pair of edges. We take out the $11th$ row and $11th$ column of vertices on the top surface as two sets of points to locally interpolate the curvature near the center of the plate in $x$ and $y$ direction respectively. The largest difference of $\nu_b$ for the same design angle pairs $\alpha$ and $\theta$ between both B.Cs applied is less than $0.5\%$.

By applying the bending strategies described above, we are able to generate deformed Miura-ori plates in simulation. See the simulation result in the below interactive Fig.\ref{3Dgeo} to understand the saddle geometry that results from  bending the Miura-ori. Readers may want to play with different toolbar options to better visualize the geometry.


\begin{thebibliography}{20}

\bibitem{Forbes1924} Wm.T.M. Forbes,  {Psyche} \textbf{31} (1924), pp.254-258. (doi:10.1155/1924/68247)

\bibitem {Kobayashi1998} H. Kobayashi, B. Kresling, and J.F.V. Vincent, T  {Proc. R. Soc. Lond. B Biol. Sci.} \textbf{265} (1998), pp.147-154. (doi:10.1098/rspb.1998.0276)

\bibitem{Kobayashi2003} H. Kobayashi,  M. Daimaruya, and  H. Fujita,   {Solid Mech. Appl.} \textbf{106} (2003), pp.207-216.

\bibitem{Lang} R. Lang, {\it Origami design secrets: mathematical methods for an ancient art}, 2nd edn (2011). A K Peters/CRC Press.

\bibitem {Miura1980} K. Miura,  {31st Cong. Intl. Astro. Fed.} \textbf{31} (1980), pp.1-10.

\bibitem {Miura1985}K. Miura and  M. Natori,  {Space Solar Power Rev.} \textbf{5} (1985), pp.345-356.

\bibitem {Elsayed2004} E.A. Elsayed and  B.B. Basily,   {Int. J. Mater. Prod. Tec.} \textbf{21} (2004), pp.217-238. (doi:10.1504/IJMPT.2004.004753)

\bibitem{Demaine} E. Demaine and J. O'Rourke, {\it Geometric folding algorithms: linkages, origami, polyhedra} (2007). Cambridge University Press.

\bibitem{Hull} T. Hull, {\it Project origami: activities for exploring mathematics} (2006). A K Peters/CRC Press.

\bibitem{Klettand2011}  Y. Klettand and K. Drechsler,   {Origami $5^th$ Intl. Meeting Origami Sci., Math. and Ed.} (2011), pp.305-322.

\bibitem{Schenk2011} M. Schenk and S. Guest,   {Origami $5^th$ Intl. Meeting Origami Sci., Math. and Ed. } (2011), pp.291-304.

\bibitem{Alessandro2008}A. Papa and S. Pellegrino,   {J. Spacecraft Rockets} \textbf{45} (2008), pp.10-18. (doi:10.2514/1.18285)

\bibitem{Whitesides1998} N. Bowden, S. Brittain, A.G. Evans, J.W. Hutchinson and G.M. Whitesides,  {Nature} \textbf{393} (1998), pp.146-149. (doi:10.1038/30193)

\bibitem {Maha2005} L. Mahadevan and S. Rica,   {Science} \textbf{307} (2005), pp.1740. (doi:10.1126/science.1105169)

\bibitem{Audoly2008} B. Audoly and A. Boudaoud,   {J Mech. Phys. Solids} \textbf{56} (2008), pp.2444-2458. (doi:10.1016/j.jmps.2008.03.001)

\bibitem{Lakes1987} R.S. Lakes,   {Science} \textbf{235} (1987), pp. 1038-1040. (doi:10.1126/science.235.4792.1038)

\bibitem{Greaves2011}  G.N. Greaves, A.L. Greer, R.S. Lakes, and T. Rouxel,   {Nature Materials} \textbf{10} (2011), pp. 823-837. (doi:10.1038/nmat3134)

\bibitem{Ventsel2001} E. Ventsel and T. Krauthammer, {\it Thin plates and shells: theory, analysis, and applications}, 1st edn (2001), CRC Press, pp.197-199.

\bibitem{Bridson2003} R. Bridson, S. Marino, and R. Fedkiw,   {ACM SIGGRAPH/Eurograph. Symp. Comp. Animation (SCA)} (2003), pp.28-36.

\bibitem{Grinspun2004} R. Burgoon, E. Grinspun, Z. Wood,   {Proc. Comp.  Applic.}, pp.180-187, 2006.

\bibitem{Barbastathis2006} W.J. Arora, A.J. Nichol, H.I. Smith, and G. Barbastathis,   {Appl. Phys. Lett.} \textbf{88} (2006). (doi: 10.1063/1.2168516)

\bibitem{Wood2010} E. Hawkes, B. An, N. Benbernou, H. Tanaka, S. Kim, E.D. Demaine, D. Rus, and R.J. Wood,  {Proc. Nat. Acad. Sci.} \textbf{107} (2010), pp.12441-12445. (doi: 10.1073/pnas.0914069107)

\bibitem{Lobkovsky1996}  A.E. Lobkovsky, Boundary layer analysis of the ridge singularity in a thin plate, {Phys. Rev. E} \textbf{53} (1996), pp.3750. (doi:10.1103/PhysRevE.53.3750)
\end{thebibliography}
\end{document}